%% file: main.tex
\documentclass{article}
\usepackage{spconf,amsmath,graphicx}
\usepackage{amssymb}
\usepackage{xcolor}
\usepackage{url}
\usepackage[hidelinks]{hyperref}
\usepackage{pifont}
\usepackage{multirow}

\definecolor{grey}{HTML}{666666}
\definecolor{purple}{HTML}{4D0E68}
\definecolor{orange}{HTML}{C94B42}
\definecolor{magenta}{HTML}{751C6C}

\usepackage{acronym}
\acrodef{ajepa}[A-JEPA]{Audio-based Joint-Embedding Predictive Architecture}
\acrodef{ema}[EMA]{Exponential Moving Average}
\acrodef{m2d}[M2D]{Masked Modeling Duo}
\acrodef{vit}[ViT]{Vision Transformer}
\acrodef{jepa}[JEPA]{Joint-Embedding Predictive Architectures}

\newcommand{\s}{\bold{s}}
\newcommand{\x}{\bold{x}}
\newcommand{\X}{\bold{X}}
\newcommand{\z}{\bold{z}}

\newcommand{\zbar}{\bold{\bar{z}}}

\newcommand{\Z}{\bold{Z}}
\newcommand{\Zbar}{\bold{\bar{Z}}}

\newcommand{\C}{\mathcal{C}}
\newcommand{\T}{\mathcal{T}}

\newcommand{\library}{AJEPA}

\renewcommand{\paragraph}[1]{\noindent\textbf{#1}\;}

\title{Investigating design choices in joint-embedding \\ predictive architectures for general audio representation learning}
\name{Alain Riou$^{1,2}$, Stefan Lattner$^2$, Gaëtan Hadjeres$^3$, Geoffroy Peeters$^1$}
\address{
	$^1$LTCI, Télécom-Paris, Institut Polytechnique de Paris, France \\
	$^2$Sony Computer Science Laboratories - Paris, France \\
	$^3$Sony AI
}
\begin{document}
%
\maketitle
\begin{abstract}
    This paper addresses the problem of self-supervised general-purpose audio representation learning. We explore the use of \ac{jepa} for this task, which consists of splitting an input mel-spectrogram into two parts (context and target), computing neural representations for each, and training the neural network to predict the target representations from the context representations.
    
    We investigate several design choices within this framework and study their influence through extensive experiments by evaluating our models on various audio classification benchmarks, including environmental sounds, speech and music downstream tasks.
    We focus notably on which part of the input data is used as context or target and show experimentally that it significantly impacts the model’s quality.
    In particular, we notice that some effective design choices in the image domain lead to poor performance on audio, thus highlighting major differences between these two modalities.
    \end{abstract}
\section{Introduction}
\label{sec:intro}



Self-supervised learning techniques are now becoming essential in training powerful models \cite{DINO,IJEPA} as they allow to create meaningful representations leveraging large quantities of unlabeled data.
These representations can then be used as inputs for various small models to solve several downstream tasks, thus preventing the need to train one huge model per task.

To learn such representations, most methods optimize a contrastive objective using inputs that share semantic information as positive pairs~\cite{CPC,SimCLR}. However, these approaches usually require many negative samples to work, making them computationally expensive.
To tackle this issue, \cite{BYOL,SimSiam} propose to use a \emph{context} and \emph{target} networks, which enables to train the model with only positive pairs artificially created using semantic-preserving data augmentations or masking.

In particular, recent methods \cite{MAE,data2vec} mask information from the data and train a Transformer \cite{Attention} to predict information about the missing part of the input from the visible one, which enforces the model to learn high-level features that capture underlying semantic information.

\begin{figure}
    \includegraphics[width=0.47\textwidth]{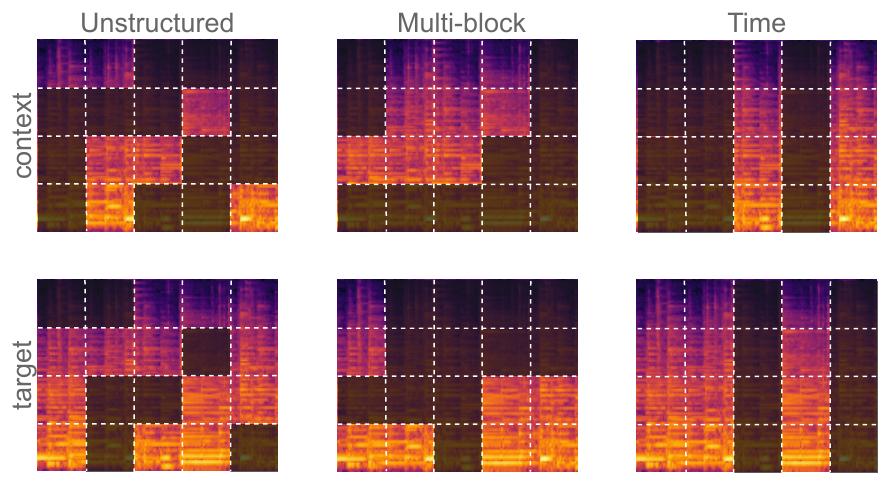}
    \caption{Investigated masking strategies. For \emph{Multi-block} and \emph{Time} strategies, target and context blocks are sampled independently, and then patches in the target blocks are eventually removed from the context one.
    }
    \label{fig:mask}
\end{figure}

Although initially proposed for computer vision applications, this method is agnostic to the data domain and has also been applied to general-purpose audio representation learning~\cite{ATST, MSMMAE, M2D}.
In particular, recent approaches manage to learn strong and general representations suited for both environmental sounds, speech and music downstream tasks by relying on those principles.
These methods induce many design choices, and we propose to study some of them in this paper. We specifically investigate the effect of the masking strategy (see Figure \ref{fig:mask}), as well as the influence of the duration of the audios provided to the model during training.

\begin{figure*}
    \centering
    \includegraphics[width=\textwidth]{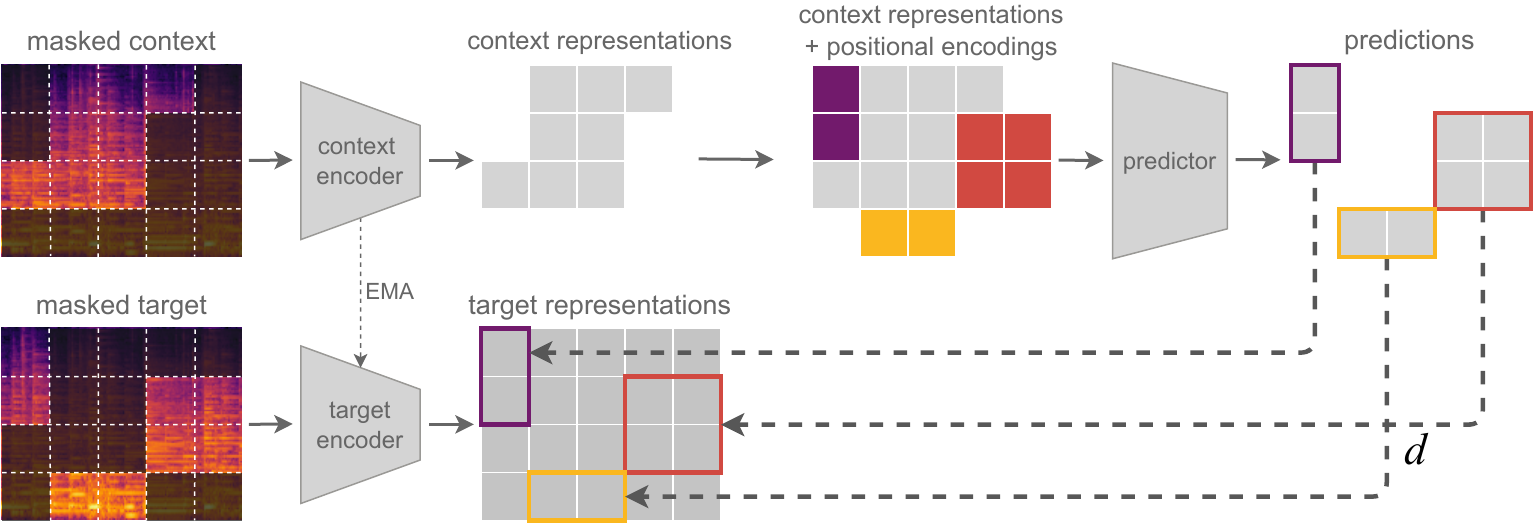}
    \caption{Overview of our framework. From an input mel-spectrogram, we first extract context and target patches via masking. These masked inputs are fed through their respective encoders to produce context and target representations. We then add positional embeddings to the context representations at the target's positions and pass the constructed sequence through a predictor, whose patch-level outputs are finally compared to the target representations.}
	\label{fig:model}
\end{figure*}

We evaluate the effectiveness of these choices by conducting extensive experiments over a wide variety of datasets and downstream tasks, thus confirming their relevance for audio representation learning.

Finally, we provide our training and evaluation procedures, along with pretrained models, for facilitating further research in general-purpose audio representation learning\footnote{\scriptsize\url{https://github.com/SonyCSLParis/audio-representations}}.

\section{Related work}
\label{sec:related}

To circumvent the lack of annotations, most self-supervised learning methods consist of applying transforms to data to create inputs that share semantic content artificially. To do so, a standard approach is to randomly apply data augmentations to inputs and train a Siamese network~\cite{Siamese} to learn similar latent representations for these transformed inputs. To prevent representation collapse (i.e., all inputs being mapped to the same latent representation), a common technique is to add a repulsive term to the model's objective by treating the elements in a batch as negative samples \cite{SimCLR} or by directly optimizing some batch-wise statistics~\cite{Wang2020a,BarlowTwins,VICReg}.
Another solution consists of adding a predictor after one branch of the Siamese network and a stop-gradient operator after the other one~\cite{BYOL,SimSiam}, which has been shown to prevent representation collapse implicitly~\cite{Tian2021}.
While initially proposed for images, all these methods have also been successfully applied to audio representation learning as well~\cite{ATST,COLA,BYOLA}.

Nevertheless, data augmentations are designed for specific modalities and may discard valuable information from the learned representations, such as color or pitch in the image and audio domains respectively.
Instead of using such handcrafted transforms, one can partially mask the input and train the model to reconstruct the masked part given the visible one, either in the domain space~\cite{MAE,MSMMAE} or in the latent space~\cite{IJEPA,data2vec,M2D}. This masking strategy is modality-agnostic, leads to general representations and is highly compatible with Transformers~\cite{Attention}, masked patches being analogous to masked tokens in Masked Language Modeling~\cite{BERT}. In particular, \ac{m2d} \cite{M2D} combines an asymmetric architecture with this masking approach to produce general-purpose audio representations.

\section{Joint-Embedding\\Predictive Architectures}
\label{sec:method}

Our method is strongly inspired from M2D~\cite{M2D} and relies on a Joint-Embedding Predictive Architecture \cite{IJEPA}. The overall idea is to train a model to generate context and target representations from an input and to be able to predict target representations from the context ones, as depicted on Figure \ref{fig:model}.

\subsection{Training method}

We first convert the input audio waveform into a log-scaled mel-spectrogram $\X$ and divide it into a sequence of $N$ non-overlapping patches $\x_1, \dots, \x_N$ over the time and frequency axes. Then, we sample two disjoint sets of indices $\C, \T \subset \{1, \dots, N\}$ and build a \emph{context} (resp. \emph{target}) sequence $\X_\C = (\x_j)_{j \in \C}$ (resp. $\X_\T = (\x_j)_{j \in \T}$), i.e. we mask $\X$ to produce two sequences of patches $\X_\C$ and $\X_\T$ (see Figure \ref{fig:model}).

After this, the context sequence $\X_\C$ is fed to a \ac{vit} $f_{\theta}$ of parameters $\theta$, called \emph{context encoder}, to produce patch-level context representations $\Z_\C = (\z_j)_{j \in \C}$ \cite{ViT}.
Analogously, target representations $\Zbar_\T = (\zbar_j)_{j \in \T}$ are computed by passing $\X_\T$ through a \emph{target encoder} $f_{\bar{\theta}}$, whose parameters $\bar{\theta}$ are updated using an \ac{ema} of those of the context encoder:
\begin{equation}
    \bar{\theta}_t = \tau_t \bar{\theta}_{t-1} + (1 - \tau_t) \theta_t
\end{equation}
where the \ac{ema} rate $\tau_t$ is linearly interpolated between $\tau_0$ and $\tau_{T}$, $T$ being the total number of training steps.

Finally, a predictor $g_\phi$ is conditioned on the context representations and the target's positions to generate a sequence $\hat{\Z} = (\hat{\z}_j)_{j \in \T}$, i.e. $\hat{\Z} = g_\phi(\Z_\C, \T)$.

Concretely, we concatenate positional encodings\footnote{As in \cite{IJEPA}, positional encodings are sinusoidal positional embeddings added to a shared learnable mask embedding.} that indicate the location of the target patches to the context representations $\Z_\C$ and pass the resulting sequence through $g_\phi$, which is also a \ac{vit}. We then select the positions corresponding to the targets in the output sequence to build $\hat{\Z}$.

The parameters $(\theta, \phi)$ of the context encoder and predictor are updated through gradient descent by minimizing the distance $\mathcal{L}$ between $\hat{\Z}$ and the target sequence $\Zbar_\T$:
\begin{equation}
    \mathcal{L} = \dfrac{1}{|\T|} \sum_{j \in \T} d(\hat{\z}_j, \zbar_j)
\end{equation}
where $d$ denotes the smoothed $L_1$ distance:
\begin{equation}
    d(\z, \z') =
    \begin{cases}
        \frac{1}{2} \| \z - \z' \|_2^2 &\text{if } \ \| \z - \z' \|_1 < 1 \\
        \| \z - \z' \|_1 - \frac{1}{2} &\text{otherwise}
    \end{cases} \\
\end{equation}
We indeed found it to work slightly better than the commonly used normalized $L_2$ distance~\cite{IJEPA,BYOL,M2D} in our preliminary experiments.

\subsection{Sampling context and target blocks}\label{sec:mask}

We investigate different masking strategies for sampling context and target sets $\C$ and $\T$ (see Figure \ref{fig:mask}). 

\paragraph{Unstructured sampling.}
The easiest and most common strategy consists in sampling the target patch indices $\T$ as a random subset of $\{1, \dots, N\}$, then using the remaining patches as context, i.e., $\C = \{1, \dots, N\} \setminus \T$, as e.g. in \cite{MAE,MSMMAE,M2D}.

\paragraph{Multi-block sampling.}
In I-JEPA~\cite{IJEPA}, the authors suggest sampling targets independently as several contiguous blocks instead of an unstructured set of patches. Then, the context is sampled as a big contiguous block from which patches already in the target set are removed.
This strategy significantly improves downstream classification performances on images (+36\% on ImageNet, see Table 6 of \cite{IJEPA} for details), we therefore investigate its effectiveness on mel-spectrograms.


\paragraph{Time sampling.}
Unlike images where elements are spatially related both vertically and horizontally, (mel-)spectrograms depict audio events spanning multiple frequency bins. To account for this distinction, we propose sampling context and target patches exclusively in the time dimension, encompassing the entire spectrum corresponding to those times.
We yet divide the frequency axis into several patches for our model to use positional encodings as indicators of the frequency range.

\input{tables/datasets}

\input{tables/sota-2}

\section{Experiments}
\label{sec:experiments}

\input{tables/mask}


\subsection{Experimental setup}

To specifically measure the influence of the design choices we study, we keep the architecture and hyperparameters unchanged across our different experiments. We choose the ones from Masked Modeling Duo (M2D)~\cite{M2D} as it is very close from our framework.

\paragraph{Model architecture.} The audio encoders $f_{\theta}$ and $f_{\bar{\theta}}$ are vanilla ViT-Base \cite{ViT} with Flash-Attention~\cite{FlashAttention}, 12 predicting heads and 12 768-dimensional layers. 
The predictor $g_{\phi}$ is a narrow \ac{vit} with 16 heads and 8 512-dimensional layers. The size of the patches $\s_j$ is fixed to $16 \times 16$.

\paragraph{Input representation.} The input audio is subsampled to 16 kHz and then converted into a log-scaled mel-spectrogram with the same settings as in \cite{M2D} (80 mel-bins ranging from 50 to 8000 Hz). Since the encoder only accepts fixed-size inputs, we pad or crop the model's inputs to 208 frames, representing each mel-spectrogram as a sequence of $(80 / 16) \times (208 / 16) = 65 = N$ patches.
We discover the choice of this parameter to be critical and study its influence in section \ref{sec:results}.

\paragraph{Pre-training details.} We train our model on the unbalanced train segment of AudioSet \cite{AudioSet}, which contains about 2 million 10-second-length audio files, during 300 epochs with a base learning of 0.0003 and AdamW \cite{DBLP:conf/iclr/LoshchilovH19} as optimizer.
All hyperparameters are kept identical to the ones in \cite{M2D}.

\paragraph{Linear evaluation details.} To evaluate how general the representations learned by our model are, we train a linear classifier on top of the \emph{frozen} pre-trained context encoder (without any masking) and evaluate it on eight downstream classification tasks, including various environmental, speech, and music datasets (see Table \ref{tab:datasets}).
Since our ViT encoder does not return one but a sequence of 768-d embeddings, we compute latent representations by concatenating them along the frequency axis, and then we average-pool along the time dimension to get a single 3840-d vector is then fed to the classifier.

\subsection{Results}\label{sec:results}



In Table \ref{tab:sota}, we present our method's performance on linear downstream classification tasks and investigate the impact of the duration of the audio samples used during training.
We observe a strong, task-dependent correlation between training sample duration and performance: tasks like environmental sounds or music genre recognition benefit from a wider context, while those emphasizing short-time semantic information (instrument, pitch or word identification) show better results with shorter segments, as observed also in \cite{Quelennec2023}.

To highlight the effect of our design choices, we compare our models to several state-of-the-art methods.
We focus on fully self-supervised methods that do not rely either on fine-tuning, ensembling or knowledge distillation, and we use the same code for evaluating all methods\footnote{\url{https://github.com/nttcslab/eval-audio-repr}}, ensuring fair comparisons.
Notably, ATST~\cite{ATST} consistently performs worse on all tasks except ESC-50, highlighting the superiority of masking over data augmentations for self-supervised audio representation learning.
Moreover, our models emerge as top performers in two tasks and second-best in another, surpassing alternative methods. Specifically, the one with a duration parameter of $d = 2.1$ s outperforms all baselines in three out of eight tasks while demanding much less GPU memory.

\subsection{Influence of the masking strategy}




In Table~\ref{tab:mask}, we investigate the influence of the various masking strategies described in section \ref{sec:mask}.
In contrast to findings in image representation learning~\cite{IJEPA}, where multi-block masking greatly increases the downstream performances by favoring local connectivity, we observe that unstructured masking significantly outperforms the alternatives across all tasks for mel-spectrograms.
This discrepancy can be attributed to the wide frequency range covered by audio events, rendering the advantages of local connectivity less relevant in this context.

Finally, we investigate the impact of masking targets only in the latent domain: the target encoder $f_{\bar{\theta}}$ can then use the entire input to compute the target representations $\Zbar_\T$, which has been found beneficial in I-JEPA~\cite{IJEPA}.
In contrast to this approach's efficacy for images, we observe in Table \ref{tab:mask} that not masking the target mel-spectrograms before passing them through the target encoder actually degrades the quality of the learned representations. Interestingly, though, with these settings, the multi-block strategy outperforms the unstructured one for a few downstream tasks, as observed for images \cite{IJEPA}.

\section{Conclusion}
\label{sec:conclusion}


In this study, we explore various design choices for \ac{jepa}s in the context of general-purpose audio representation learning. By examining the impact of masking strategies and target masking, we empirically demonstrate that optimal design choices differ between audio and image domains. Specifically, the effective multi-block masking strategy introduced in \cite{IJEPA}, advantageous for local connectivity in natural images, proves less suitable for mel-spectrograms.

Moreover, our experiments emphasize the importance of the duration of training audio samples, showing varied effects depending on the downstream task. Notably, longer context negatively impacts representations in certain tasks.
This observation reveals the need for further exploration to enhance the adaptability of audio representation learning architectures to different temporal scales.

Overall, ViT architectures exhibit notable performance across various audio downstream tasks, and our findings highlight the relevance of continuing research in this direction to improve general-purpose audio representation learning.

\section{Acknowledgments}

This work has been funded by the ANRT CIFRE convention n°2021/1537 and Sony France. This work was granted access to the HPC/AI resources of IDRIS under the allocation 2022-AD011013842 made by GENCI. We would like to thank the reviewers and meta-reviewer for their valuable and insightful comments.

\bibliographystyle{IEEEbib}
{\small
\bibliography{\library}
}

\end{document}

%% file: tables/datasets.tex
\begin{table}
    \centering
    \caption{Downstream tasks used for linear evaluation.}
    \label{tab:datasets}
    \small
            {\small
    \begin{tabular}{lcl}
        \hline
        Dataset & classes & Task \\
        \hline
        ESC-50 \cite{piczak2015dataset}                     & 50 & Env. sound classification       \\
        UrbanSound8K (US8K) \cite{Salamon:UrbanSound:ACMMM:14}         & 10 & Urban sound classification \\
        Speech Commands V2 & \multirow{2}{*}{35} & \multirow{2}{*}{Word classification}         \\
         (SPCV2) \cite{SPCV2}   &  &  \\
        VoxCeleb1 (VC1) \cite{Nagrani17}              & 1251 & Speaker identification     \\
        CREMA-D (CRM-D) \cite{DBLP:journals/taffco/CaoCKGNV14}             &  6 & Emotion recognition          \\
        GTZAN \cite{tzanetakis_essl_cook_2001}                      & 10 & Music genre recognition      \\
        NSynth \cite{pmlr-v70-engel17a}                    & 11 & Instrument classification    \\
        Surge \cite{DBLP:conf/dafx/TurianSTMH21}                      & 88 & Pitch classification         \\
        \hline
    \end{tabular}
    }
\end{table}

%% file: tables/sota-2.tex
\begin{table*}
	\centering
	\caption{
        Linear classification accuracies on several downstream tasks. We report our results when training the model on audio segments of different durations $d$ and also compare with other methods.
        Best results are bold and second-to-best are underlined.
    }
	\label{tab:sota}
         {\small
	\begin{tabular}{lccccccccccccc}
		\hline
         & & \multicolumn{2}{c}{Environmental} & & \multicolumn{3}{c}{Speech} & & \multicolumn{3}{c}{Music} \\
        \cline{3-4} \cline{6-8} \cline{10-12}
		Model & &
		ESC-50 &
        US8K & &
		SPCV2 &
        VC1 &
		CRM-D & &
		GTZAN &
        NSynth &
        Surge \\
		\hline
        Ours ($d = 2.1$ s) &  &
        89.3 & 87.2 &  & 94.9 & 60.8 & \underline{72.2} &  & 82.1 & \textbf{76.8} & \textbf{42.8} \\
        Ours ($d = 3.2$ s) &  &
        89.8 & 86.5 &  & 94.6 & 61.7 & \textbf{72.6} &  & \underline{84.3} & 75.6 & 42.0 \\
        Ours ($d = 6.4$ s) &  &
        \underline{90.0} & \textbf{87.7} &  & 92.8 & 63.6 & 70.8 &  & \textbf{86.9} & 74.9 & 41.1 \\
		\hline
        Wav2Vec2.0 \cite{Wav2Vec2} & &
        57.6 & 66.9 &  & \textbf{96.6} & 40.9 & 65.5 &  & 57.8 & 56.6 & 15.2 \\
        ATST Base \cite{ATST} & &
        \textbf{92.9} & 84.1 &  & 95.1 & 72.0 & 68.6 &  & 76.4 & 75.6 & 37.7 \\
        MSM-MAE \cite{MSMMAE} &  &
        88.6 & 86.3 &  & 94.5 & \underline{72.2} & 70.2 &  & 78.4 & \underline{75.9} & \underline{42.5} \\
        M2D \cite{M2D} & &
        89.7 & \underline{87.6} &  & \underline{95.4} & \textbf{73.1} & 71.7 &  & 83.3 & 75.3 & 41.0 \\
		\hline
	\end{tabular}
 }
\end{table*}

%% file: tables/mask.tex
\begin{table*}
	\centering
	\caption{Influence of the masking strategy and of whether targets are masked in the audio or latent domain ($d = 2.1$ s).}
	\label{tab:mask}
        {\small
    \begin{tabular}{lcccccccccc}
        \hline
         & \multicolumn{2}{c}{Environmental} & & \multicolumn{3}{c}{Speech} & & \multicolumn{3}{c}{Music} \\
        \cline{2-3} \cline{5-7} \cline{9-11}
		Masking strategy &
		ESC-50 &
		US8K &  &
		SPCV2 &
		VC1 &
		CRM-D &  &
		GTZAN &
		NSynth &
		Surge \\
		\hline
        \emph{Masking target in the audio domain} \\
		Unstructured &
        \textbf{89.3} & \textbf{87.2} &  & \textbf{94.9} & \textbf{60.8} & \textbf{72.2} &  & \textbf{82.1} & \textbf{76.8} & \textbf{42.8} \\
        Multi-block &
        88.6 & 86.4 &  & 92.1 & 54.7 & 67.6 &  & 81.6 & 75.2 & \textbf{42.8} \\
        Time &
        78.1 & 77.1 &  & 86.7 & 52.3 & 62.5 &  & 79.2 & 71.7 & 33.0 \\
        \emph{Masking target in the latent domain} \\
		Unstructured &
        85.1 & 84.1 &  & 92.3 & 56.9 & 66.4 &  & 80.1 & 76.5 & 40.8 \\
        Multi-block &
        88.4 & 79.6 &  & 89.4 & 54.3 & 66.3 &  & 83.1 & 72.8 & 35.2 \\
        Time &
        82.3 & 81.0 &  & 85.2 & 30.6 & 60.9 &  & 75.6 & 73.7 & 37.7 \\
		\hline
	\end{tabular}
 }
\end{table*}

%% file: main.bbl
\begin{thebibliography}{10}

\bibitem{DINO}
Mathilde Caron, Hugo Touvron, Ishan Misra, Herv{\'{e}} J{\'{e}}gou, Julien
  Mairal, Piotr Bojanowski, and Armand Joulin,
\newblock ``{Emerging Properties in Self-Supervised Vision Transformers},''
\newblock in {\em ICCV}, apr 2021, pp. 9630--9640.

\bibitem{IJEPA}
Mahmoud Assran, Quentin Duval, Ishan Misra, Piotr Bojanowski, Pascal Vincent,
  Michael Rabbat, Yann LeCun, and Nicolas Ballas,
\newblock ``{Self-Supervised Learning from Images with a Joint-Embedding
  Predictive Architecture},''
\newblock jan 2023.

\bibitem{CPC}
A{\"{a}}ron van~den Oord, Yazhe Li, and Oriol Vinyals,
\newblock ``{Representation Learning with Contrastive Predictive Coding},''
\newblock jul 2018.

\bibitem{SimCLR}
Ting Chen, Simon Kornblith, Mohammad Norouzi, and Geoffrey Hinton,
\newblock ``{A simple framework for contrastive learning of visual
  representations},''
\newblock in {\em ICML}, feb 2020, pp. 1575--1585.

\bibitem{BYOL}
Jean~Bastien Grill, Florian Strub, Florent Altch{\'{e}}, Corentin Tallec,
  Pierre~H. Richemond, Elena Buchatskaya, Carl Doersch, Bernardo~Avila Pires,
  Zhaohan~Daniel Guo, Mohammad~Gheshlaghi Azar, Bilal Piot, Koray Kavukcuoglu,
  R{\'{e}}mi Munos, and Michal Valko,
\newblock ``{Bootstrap your own latent a new approach to self-supervised
  learning},''
\newblock in {\em Advances in Neural Information Processing Systems}, jun 2020.

\bibitem{SimSiam}
Xinlei Chen and Kaiming He,
\newblock ``{Exploring simple Siamese representation learning},''
\newblock in {\em CVPR}, nov 2021, pp. 15745--15753.

\bibitem{MAE}
Kaiming He, Xinlei Chen, Saining Xie, Yanghao Li, Piotr Dollar, and Ross
  Girshick,
\newblock ``{Masked Autoencoders Are Scalable Vision Learners},''
\newblock in {\em CVPR}, nov 2022, pp. 15979--15988.

\bibitem{data2vec}
Alexei Baevski, Wei-Ning Hsu, Qiantong Xu, Arun Babu, Jiatao Gu, and Michael
  Auli,
\newblock ``{data2vec: A General Framework for Self-supervised Learning in
  Speech, Vision and Language},''
\newblock feb 2022.

\bibitem{Attention}
Ashish Vaswani, Noam Shazeer, Niki Parmar, Jakob Uszkoreit, Llion Jones,
  Aidan~N. Gomez, {\L}ukasz Kaiser, and Illia Polosukhin,
\newblock ``{Attention is all you need},''
\newblock in {\em Advances in Neural Information Processing Systems}, jun 2017,
  pp. 5999--6009.

\bibitem{ATST}
Xian Li and Xiaofei Li,
\newblock ``{ATST: Audio Representation Learning with Teacher-Student
  Transformer},''
\newblock in {\em INTERSPEECH}, apr 2022, pp. 4172--4176.

\bibitem{MSMMAE}
Daisuke Niizumi, Daiki Takeuchi, Yasunori Ohishi, Noboru Harada, and Kunio
  Kashino,
\newblock ``{Masked Spectrogram Modeling using Masked Autoencoders for Learning
  General-purpose Audio Representation},''
\newblock {\em PMLR}, pp. 1--24, 2022.

\bibitem{M2D}
Daisuke Niizumi, Daiki Takeuchi, Yasunori Ohishi, Noboru Harada, and Kunio
  Kashino,
\newblock ``{Masked Modeling Duo: Learning Representations by Encouraging Both
  Networks to Model the Input},''
\newblock oct 2023, pp. 1--5.

\bibitem{Siamese}
Raia Hadsell, Sumit Chopra, and Yann LeCun,
\newblock ``{Dimensionality reduction by learning an invariant mapping},''
\newblock {\em CVPR}, pp. 1735--1742, 2006.

\bibitem{Wang2020a}
Tongzhou Wang and Phillip Isola,
\newblock ``{Understanding contrastive representation learning through
  alignment and uniformity on the hypersphere},''
\newblock in {\em ICML}, may 2020, pp. 9871--9881.

\bibitem{BarlowTwins}
Jure Zbontar, Li~Jing, Ishan Misra, Yann LeCun, and St{\'{e}}phane Deny,
\newblock ``{Barlow Twins: Self-Supervised Learning via Redundancy
  Reduction},''
\newblock {\em ICML}, mar 2021.

\bibitem{VICReg}
Adrien Bardes, Jean Ponce, and Yann LeCun,
\newblock ``{VICReg: Variance-Invariance-Covariance Regularization for
  Self-Supervised Learning},''
\newblock in {\em ICLR}, may 2022.

\bibitem{Tian2021}
Yuandong Tian, Xinlei Chen, and Surya Ganguli,
\newblock ``{Understanding self-supervised Learning Dynamics without
  Contrastive Pairs},''
\newblock feb 2021.

\bibitem{COLA}
Aaqib Saeed, David Grangier, and Neil Zeghidour,
\newblock ``{Contrastive learning of general-purpose audio representations},''
\newblock in {\em ICASSP}. oct 2021, pp. 3875--3879, Institute of Electrical
  and Electronics Engineers Inc.

\bibitem{BYOLA}
Daisuke Niizumi, Daiki Takeuchi, Yasunori Ohishi, Noboru Harada, and Kunio
  Kashino,
\newblock ``{BYOL for Audio: Exploring Pre-Trained General-Purpose Audio
  Representations},''
\newblock {\em TASLP}, pp. 1--15, apr 2022.

\bibitem{BERT}
Jacob Devlin, Ming~Wei Chang, Kenton Lee, and Kristina Toutanova,
\newblock ``{BERT: Pre-training of Deep Bidirectional Transformers for Language
  Understanding},''
\newblock {\em NAACL HLT 2019 - 2019 Conference of the North American Chapter
  of the Association for Computational Linguistics: Human Language Technologies
  - Proceedings of the Conference}, pp. 4171--4186, oct 2018.

\bibitem{ViT}
Alexey Dosovitskiy, Lucas Beyer, Alexander Kolesnikov, Dirk Weissenborn,
  Xiaohua Zhai, Thomas Unterthiner, Mostafa Dehghani, Matthias Minderer, Georg
  Heigold, Sylvain Gelly, Jakob Uszkoreit, and Neil Houlsby,
\newblock ``An image is worth 16x16 words: Transformers for image recognition
  at scale,''
\newblock in {\em {ICLR}}, 2021.

\bibitem{piczak2015dataset}
Karol~J. Piczak,
\newblock ``{ESC}: {Dataset} for {Environmental Sound Classification},''
\newblock in {\em Proceedings of the 23rd {Annual ACM Conference} on
  {Multimedia}}. 2025, pp. 1015--1018, {ACM Press}.

\bibitem{Salamon:UrbanSound:ACMMM:14}
J.~Salamon, C.~Jacoby, and J.~P. Bello,
\newblock ``A dataset and taxonomy for urban sound research,''
\newblock in {\em 22nd {ACM} International Conference on Multimedia
  (ACM-MM'14)}, Nov. 2014, pp. 1041--1044.

\bibitem{SPCV2}
Pete Warden,
\newblock ``Speech commands: {A} dataset for limited-vocabulary speech
  recognition,''
\newblock {\em CoRR}, 2018.

\bibitem{Nagrani17}
A.~Nagrani, J.~S. Chung, and A.~Zisserman,
\newblock ``Voxceleb: a large-scale speaker identification dataset,''
\newblock in {\em INTERSPEECH}, 2017.

\bibitem{DBLP:journals/taffco/CaoCKGNV14}
Houwei Cao, David~G. Cooper, Michael~K. Keutmann, Ruben~C. Gur, Ani Nenkova,
  and Ragini Verma,
\newblock ``{CREMA-D:} crowd-sourced emotional multimodal actors dataset,''
\newblock {\em {IEEE} Trans. Affect. Comput.}, , no. 4, pp. 377--390, 2014.

\bibitem{tzanetakis_essl_cook_2001}
George Tzanetakis, Georg Essl, and Perry Cook,
\newblock ``Automatic musical genre classification of audio signals,''
\newblock in {\em ISMIR}, 2001.

\bibitem{pmlr-v70-engel17a}
Jesse Engel, Cinjon Resnick, Adam Roberts, Sander Dieleman, Mohammad Norouzi,
  Douglas Eck, and Karen Simonyan,
\newblock ``Neural audio synthesis of musical notes with {W}ave{N}et
  autoencoders,''
\newblock in {\em ICML}, 06--11 Aug 2017, pp. 1068--1077.

\bibitem{DBLP:conf/dafx/TurianSTMH21}
Joseph Turian, Jordie Shier, George Tzanetakis, Kirk McNally, and Max Henry,
\newblock ``One billion audio sounds from gpu-enabled modular synthesis,''
\newblock in {\em DAFx}, 2021, pp. 222--229.

\bibitem{Wav2Vec2}
Alexei Baevski, Henry Zhou, Abdelrahman Mohamed, and Michael Auli,
\newblock ``{wav2vec 2.0: A framework for self-supervised learning of speech
  representations},''
\newblock in {\em Advances in Neural Information Processing Systems}, jun 2020.

\bibitem{FlashAttention}
Tri Dao, Daniel~Y. Fu, Stefano Ermon, Atri Rudra, and Christopher Ré,
\newblock ``Flashattention: Fast and memory-efficient exact attention with
  io-awareness,'' 2022.

\bibitem{AudioSet}
Jort~F. Gemmeke, Daniel~P.W. Ellis, Dylan Freedman, Aren Jansen, Wade Lawrence,
  R.~Channing Moore, Manoj Plakal, and Marvin Ritter,
\newblock ``{Audio Set: An ontology and human-labeled dataset for audio
  events},''
\newblock {\em ICASSP}, pp. 776--780, 2017.

\bibitem{DBLP:conf/iclr/LoshchilovH19}
Ilya Loshchilov and Frank Hutter,
\newblock ``Decoupled weight decay regularization,''
\newblock in {\em {ICLR}}, 2019.

\bibitem{Quelennec2023}
Aurian Quélennec, Michel Olvera, Geoffroy Peeters, and Slim Essid,
\newblock ``On the choice of the optimal temporal support for audio
  classification with pre-trained embeddings,''
\newblock in {\em ICASSP}, 2024, pp. 976--980.

\end{thebibliography}
